\documentstyle[aps,twocolumn,psfig]{revtex}  

\begin{document}
\draft                       

\title{Classical molecular dynamics simulations \\ 
       of amorphous silica surfaces}

\author{M. Rarivomanantsoa, P. Jund, and R. Jullien}

\address{ Laboratoire des Verres - Universit\'e Montpellier 2 \\
        \it Place E. Bataillon Case 069, 34095 Montpellier France}


\maketitle


\begin{abstract}
We have adapted classical molecular dynamics to study the structural 
and dynamical properties of amorphous silica surfaces. 
Concerning the structure, the density profile exhibits oscillations 
perpendicularly to the surface as observed in liquid metal
surfaces and the pair correlation functions as well as the angle
distributions show features (absent in the interior of the films) that
can be attributed to the presence of 2-fold rings which are perpendicular 
to the surface. From the mean-squared displacement of the non bridging 
oxygen atoms we find that in the interior region they move perpendicular
to the surface while they move parallel to it in the surface region.
\end{abstract}
 


\narrowtext

\section{introduction}
Most studies on surfaces have been focused on their electronic aspect. 
The goal was usually to analyze materials used in electronic devices 
which characteristics depend on the surface properties. 
As a consequence, crystalline surfaces are currently well
known and a lot of attention has been devoted to the doping and transport 
properties of these surfaces as well as to the surface
reconstruction. In the last years, many 
studies, theoretical \cite{celestini1} or experimental \cite{magnussen} 
have also been devoted to the analysis of liquid 
metal surfaces. In particular, the existence of an atomic layering in 
the density profile has been found. A great interest in understanding 
amorphous surfaces exists \cite{pantano}, because of their participation in 
the glass formation, and therefore several numerical studies are now 
available on this subject \cite{garofalini,kilian,wk}. 
Pioneering work in using classical potentials to describe
amorphous silica surfaces is due to Garofalini et {\em al}. who studied
mainly the presence of defects (non bridging oxygens (NBOs), small-sized 
rings..) at the surface \cite{garofalini}. These defect sites are the most 
chemically reactive ones but experimental studies have, so far, not clearly 
shown their existence \cite{pantano,nbo}. An alternative way to learn more
about the surface is to study the chemical reactions taking place on it
\cite{chem} and in particular the interaction between water and 
amorphous SiO$_2$ surfaces has been the topic of both experimental \cite{watex,watex2}
and numerical studies \cite{watsim,baka}. 
 In most of the numerical studies available in the literature, the total 
number of atoms did not exceed 800 atoms and only a fraction of these atoms 
were considered as ``mobile'' (i.e. participating in the formation of the 
surface). (It should be noted that very recently (in fact during the process 
of writing the present article) a molecular-dynamics study of silica 
surfaces has been proposed involving several thousands of atoms \cite{wk}).
Moreover the long-range Coulomb forces were either cut-off 
\cite{garofalini} or treated in a {\em pseudo} 2D geometry \cite{method}, 
not taking into account exactly the reduced dimensionality of free silica 
surfaces. A recent study of the influence of long-range forces on the 
structure and dynamics of a model silica glass has shown that over-damped 
Coulomb interactions can alter the quality of the results \cite{jpcm}. 

In the present work the molecular dynamics force calculation scheme has 
been adapted to the specific 2D geometry of samples with free surfaces. 
Our aim was mainly to study in detail the structural properties of amorphous 
silica surfaces with respect to what is already known for the corresponding 
bulk samples and to test the quality of the widely used ``BKS'' potential 
proposed by van Beest {\em et al.} \cite{vanbeest}. This classical pair potential 
describes quite well the structural \cite{jj99,vollmayr}, and 
vibrational \cite{taraskin}, as well as relaxational \cite{kob} and 
thermal \cite{jjpr} properties of both, supercooled viscous liquid and 
glassy bulk silica samples: it is therefore justified to address its 
effectiveness in the case of ``bi-dimensional'' silica systems even more 
so since in a recent study based on {\em ab initio} simulations, 
Ceresoli et {\em al}. have put into question the structural description of 
the surfaces obtained with the BKS potential \cite{bernas}. In addition to 
the work already proposed by Roder {\em et al.} \cite{wk} who used the same 
potential but aimed their study mainly on silica {\em clusters}, we show that 
the effect of the surface is firstly reflected in the density profile which 
exhibits stratification, explained as a tendency to have more short range 
order at the surface. 
We then analyze in detail the ring size distribution as well as the
orientation of the rings with respect to the surface, and show that the number 
of two-fold rings (whose signature is visible in the radial pair distribution
functions and the angle distributions) is coherent with the experimental 
estimates and that they are positioned perpendicularly to the surface. 
By analyzing the influence of the annealing time on this effect, we show
that it is a true surface reconstruction and not an artifact due to 
out-of equilibrium phenomena.
Finally, in order to elucidate the diffusion process at the surface, 
we present the mean square displacement of the bridging and non-bridging 
oxygen atoms. We find no signature of a liquid like behavior at the surface but
the dynamics of the NBOs seems to be anisotropic.

\section{simulations}

We have performed classical molecular dynamics simulations starting with cubic 
samples of edge length L=35.80 \AA, containing 3000 particles. These 
samples were used previously to study the properties of bulk silica
therefore their mass density is $\rho \simeq  2.18$ g/cm$^3$, which is very 
close to the experimental value of 2.20 g/cm$^3$ \cite{mazur}, and the 
periodic 
boundary conditions were assumed in the three space directions. The potential 
used to describe the interactions between the atoms is the two-body ``BKS'' potential. After the use of this potential to study densified silica samples 
\cite{jcp}, we apply it here to study silica surfaces assuming this potential 
is still valid in the case of SiO$_2$ free surfaces. Of course this 
description can never be as accurate as an {\em ab initio} study, nevertheless we can simulate much larger samples and typical structural features 
should not depend on the details of the potential.  The classical equations 
of motion are resolved by using the Velocity-Verlet algorithm with a 
timestep $\Delta t=$ 0.7 fs. The initial configuration is given by a bulk 
sample of amorphous silica at $T = 0$ K. This configuration has been 
obtained by quenching a well equilibrated liquid sample around 7000 K with 
a quench rate of $2.3\times10^{14}$ K/s. The free surfaces are created by 
breaking the periodic boundary conditions along the $z$-direction, 
normal to the surface, thus creating two free surfaces located at $L/2$ and
$-L/2$. Doing so, the Ewald summation for the calculation of the long range 
forces has to be modified to take into account the loss of periodicity in 
the $z$-direction. 
There exists no standard method since several possibilities based on
either a modified Ewald summation \cite{method,ewal2} or multipole 
expansions \cite{multip} are proposed in the literature.
A purely two-dimensional method exists \cite{leeuw1} which nevertheless 
necessitates the adjunction of charged plates.
Here we have chosen a strictly two-dimensional {\em approximate} technique 
which has the advantage to be very simple \cite{ahermouch}. In the three 
dimensional Ewald summation technique, one considers 
all the point charges surrounded by a spherical distribution of charges 
that has same magnitude but opposite sign 
\cite{allen}. Then, the potential is separated into two terms. 
The first term $V_c$ corresponds to the potential due to the 
point charge distribution minus that of the screening spherical 
charge distribution and is quite rapidly convergent in real
space. The second one $V_{\gamma}$ is the potential created by the 
spherical distribution and is rapidly convergent in the
reciprocal space. Since the real space term is a short range potential 
with a spherical symmetry, there is no need to change 
it for the two-dimensional summation. But, of course    
the reciprocal space potential has to be modified. While for the 
3d Ewald summation the expression for the reciprocal space term of the 
potential is:
\begin{eqnarray}
V_{\gamma} = \frac{2\pi}{L^3} \sum_{\vec{k}\neq 0}\frac{1}{k^2}e^{-\frac{k^2}{4\kappa^2}}
                 \left|\sum_{i}q_ie^{-i\vec{k}.\vec{r_i}}\right|^2
\end{eqnarray}
with
\begin{eqnarray}
\vec{k}=n_x\frac{2\pi}{L}\vec{e_x}+n_y\frac{2\pi}{L}\vec{e_y}+n_z\frac{2\pi}{L}\vec{e_z},
\end{eqnarray}
In our approximate two-dimensional method it becomes:
\begin{eqnarray}
V_{\gamma} = \frac{\pi}{L^2} \sum_{\vec{k}\neq 0}\frac{e^{-\frac{k^2}{4\kappa^2}}}{k}\left|\sum_{i}q_ie^{-i\vec{k}.\vec{r_i}}\right|^2
\end{eqnarray}
with
\begin{eqnarray}
\vec{k}=n_x\frac{2\pi}{L}\vec{e_x}+n_y\frac{2\pi}{L}\vec{e_y}.
\end{eqnarray}
Here $\vec{k}$ is a two-dimensional vector and the summation runs 
over the two integers $n_x$ and $n_y$.
Although formula (3) looks like a straightforward two-dimensional 
transcription of formula (1) it is only approximate. The expression would 
have been exact if, instead of a Coulomb potential with
a spherical symmetry one would have considered a cylindrical potential. In 
particular, one practical consequence is that the k-contribution 
to the $z$-component of the force is zero, since the scalar product 
$\vec{k}.\vec{r_i}$ is independent of the third coordinate $z_i$ of 
$\vec{r_i}$.
We have used the characteristic constant $\kappa=5.0/L$, 
where $L$ is the cubic box size, and considered 49 k-vectors in 
reciprocal space to insure a relative error smaller than $2\times 10^{-5}$ 
for the potential energy. The removal of the periodic boundary conditions 
in the $z$-direction breaks the bonds between the
atoms at the ``bottom'' and those at the ``top'' of the film. This 
will lead to bond rearrangements near each free
surface and of course increase the temperature of the system. 
It is also worth noticing that we let the position of the atoms adjust 
freely in the $z$-direction and therefore we are not {\em stricto sensu} 
in a microcanonical ensemble.
To justify {\it a posteriori} the approximations and check the results 
we compared this method with the one in which 
the box length in the $z$-direction, L$_z$, is artificially increased in order
to simulate a pseudo 2D system. We find that the two techniques give 
similar results when L$_z$ is five times larger than the box length in 
the other 2 directions which is in contrast with the study of Bakaev et 
{\em al}. \cite{baka} in which L$_z$ was taken only twice as large as 
in the $x$ and $y$ directions. In any case the method used here is 
computationally much more efficient since the number of k-points included
in the Ewald summation has not to be increased. It is also faster than the 
more frequently used technique proposed in Ref. \cite{method} which is not
a strictly 2D technique and which is computationally very costly.\\
The heating of the slab mentioned earlier will tend to evaporate some 
oxygen atoms located at the surface, and to avoid this vaporization, it 
is necessary to control the temperature of the slab. This has been done 
at 1000K by rescaling the velocities during 30000 time steps. This 
temperature is below the simulated glass transition temperature as will be 
shown in a forthcoming study. Then the system was allowed to relax 30000 
supplemental steps and finally we performed simulations 
during 60000 supplemental steps to collect the results, the system 
temperature remaining almost constant around 1000 K in this time range. 
Such a simulation represents 45 days of computer time on the latest IBM-SP2 
processor and in order to improve the statistics of the data we 
averaged our results over ten independent liquid samples.
 
\section{results}

In figure 1 we have reported the averaged density profile of our samples 
in the $z$-direction. Each point (filled circle) of the figure corresponds 
to the density $\rho (z)$ calculated within a slab parallel to the surface 
of area $L\times L$ and of width $\Delta z = 0.4475$ \AA\ and reported as a 
function of the mean slab position $z$. We have averaged the results
over the two slabs located at $z$ and $-z$.
For comparison, we have reported the profile in the $x$-direction calculated
within a slab of same width $\Delta x = 0.4475$ \AA\ but of area $L\times L_z$
with $L_z = 26.85$ \AA, taken sufficiently smaller than $L$ to avoid edge
effects due to the presence of the free surface in the $z-$direction.
In both cases, we get an average bulk density of about $2.3$ g/cm$^3$, 
significantly larger than the density $\rho=2.18$ g/cm$^3$
of the original cubic box with periodic boundary conditions.
This increase of the density is due to the relaxation of the system (the 
system contracts along the $z$ direction) when 
creating the free surfaces  since it has been shown that the 
equilibrium density of the BKS potential is larger than the 
experimental one \cite{density} (this means that the initial sample with 
PBC had a negative pressure). This shows that creating free surfaces is 
a physical way to obtain the equilibrium density and thus permits also to 
validate a posteriori our {\em modus operandi}. For the profile in the 
$z$-direction, one can use a hyperbolic tangent fit derived from the
Van der Waals theory of surface tension \cite{chapela}. 
Although this formula is generally used to study the thermodynamic
properties of simple liquids, some characteristics of the density profile 
of our amorphous samples can be determined. This
fit uses three parameters, $\rho_0$ the mass density in the underlying bulk, 
$z_0$ the position of the surface and $d$ the 
surface thickness:
\begin{eqnarray}
\rho (z) = \frac{\rho_0}{2} \left(1-\tanh \left(\frac{2(z-z_0)}{d} \right)\right).
\end{eqnarray}
In fact, this formula fits very well the vapor-liquid interfaces 
which exhibit symmetric density profiles \cite{chapela}. But here 
it does not  take into account the asymmetric density maximum
near the free surfaces (see figure~1) and  therefore the
density profile is not very well fitted near this maximum. As
a consequence, if the parameter $\rho_0$ ($\rho_0=2.36 g/cm^3$)
gives a good quantitative value for the bulk density, the parameter
$d$ ($d=1.18$ \AA\ ) gives only a rough (underestimated) value of the actual 
surface thickness. 
As observed for liquid surfaces \cite{iarlori}, 
we have an asymmetric peak with a broad tail in the density profile at the
surface and a layering behavior. Our goal is to identify the participation 
of the stratification in the oscillations observed in the density profile 
along the $z$-direction. However when comparing the density profiles
in the $z$ and $x$ directions in figure 1, one can hardly see a difference
in the oscillations around the plateau. One way to address this point is 
to compute the standard deviation of these data over different samples. 
If the oscillations are only due to statistical 
fluctuations, the standard deviation $\sigma$ of the density should  behave 
like $N^{-1/2}_S$, where $N_S$ is the number of statistical independent 
samples over which the density profiles are averaged. Along the $x$ and 
$y$ axis, we find that the standard deviations behave like $N_S^{-1/2}$ while for the $z$-direction they are quite independent on $N_S$.
Therefore the  oscillations along the $z$-direction 
are most likely a signature of a layer formation  whereas the 
oscillations observed in the $x$-direction are only due to statistical density fluctuations. This stratification is likely to be enhanced at the free 
surfaces and to extend in the underlying bulk. Computing the radial pair 
distribution for atoms within a sphere of radius $R = L/4$, whose center 
is the center of the the initial cubic box, we found that it is similar to 
that observed for bulk amorphous silica samples. This is an evidence that 
the layering behavior is not due to crystallization. The inset of Fig. 1, 
exhibits the contribution of each type of atom to the mass density. Since 
the stoichiometry is respected, the quantity $\Delta n$~=~2[Si]~-~[O] is 
oscillating around a zero mean within the 
underlying bulk. Near the surface, the apparition of two peaks reflects 
some stoichiometry breaking. The silicon atom excess under the surface and 
the oxygen atom excess at the surface are respectively represented by the 
positive and the negative peak. This surface segregation appears not to be 
model dependent since it has been found also by Roder {\em et al.} \cite{wk} 
as well as by Litton {\em et al.} \cite{litton} with another potential.\\
In figure 2 we present the radial pair distribution functions (RPDFs) for the
different pairs of atoms in order to investigate the structure of our samples.
The space has been divided into two regions, the ``surface'' layer 
which extends 6 \AA\ below the $z$ position of the outermost atom, and
the ``interior'' which represents the rest of the sample. The width of the
surface layer has been chosen to encompass the broad peak in the density
profile visible in figure 1 and is coherent with previous studies \cite{wk}.
Also shown in figure 2 are the RPDFs obtained in a bulk silica sample.
As a general trend, the RPDFs in the interior are very similar to the
ones obtained in the bulk. Nevertheless differences can be seen especially
in the O-O RPDF, where a slight shift towards smaller distances can be seen. 
This is a consequence of the relaxation towards a higher equilibrium 
density. The major difference between the surface and the interior can be
seen in the Si-Si RPDF, since a small shoulder appears around 2.5 \AA\ .
This shoulder which can be attributed to 2-membered rings as we will see
below, is absent in the interior or bulk curve since this structural unit
is not present in the bulk \cite{jpcm,vollmayr,rino}. In figure 3 we 
show the distribution functions of the angle between an oxygen (silicon) 
atom and its 2 silicon (oxygen ) neighbors (two atoms are neighbors if
their separation is smaller than the location of the first minimum in the
corresponding RPDF) where we distinguish again the surface, the interior and
the results obtained for bulk silica (this is a further test of the 
validity of the BKS potential which does not contain 3-body terms).
Concerning the O-Si-O angle distribution, there is almost no difference
between the interior and the bulk. The distribution at the surface is 
slightly broader and shows a second peak around 80$^\circ$ which is absent in the
bulk. Concerning the Si-O-Si angle distribution we firstly note a slight
shift of the whole distribution towards smaller angles compared to
the bulk distribution which is again a consequence of the density relaxation.
Next we note a strong shift of the surface distribution towards smaller
angles by more than 10$^\circ$ which is coherent with the peak in the density
profile observed in figure 1 and an important second peak around 100$^\circ$
which is absent in the bulk and interior curves. 
The existence of the additional peaks in the surface angle distributions is
in agreement with previous studies \cite{garofalini,wk}. The location of
these peaks is different from the angles O-Si-O and Si-O-Si expected in 2 
membered rings \cite{dubois}: 94 and 86$^\circ$ respectively. Here we find 80 and
100$^\circ$ which indicates a weakness of the pair potential on this particular
point since in a recent {\em ab initio} study on silica clusters Lopez 
{\em et al.} found 91 and 89$^\circ$ respectively \cite{lopez}. \\
So far we only have the 
signature of the presence of 2-membered rings at the
surface therefore it is justified to address the influence of the free 
surfaces on the variation of the ring size distribution.  
Such quantity is of interest since it can be extracted from infrared and 
Raman spectroscopy. In particular the highly strained 2-membered rings 
result in infrared-active Si-O stretching modes at 888 and 908 cm$^{-1}$ \cite{infra1,infra2}. The ring size distribution has been determined using the 
algorithm described in \cite{kob}, where a ring is defined as the smallest 
loop starting from one oxygen atom O$_1$ nearest neighbor of a given
Si atom and ending on O$_2$ another of its nearest neighbor. 
Then the size of a ring is given by the number of its 
constitutive Si-O segments. To get the distribution of each ring 
size along the $z$ axis in the 2 regions (interior and surface) we compute 
the probability $P_n$ for a given Si atom, whose coordinate along the normal 
direction to the surface is in a given region, to be a member of a 
$n$-fold ring. These results are reported in the top part of figure~4 
and compared to the bulk distribution. We calculated the ring size distribution
in the surface layer on the initial 2D samples (not relaxed) and on the 2D 
samples annealed for 42ps (relaxed) in order to see the influence of the
annealing time on the results. Firstly it should be noted that the
distributions in the interior region (relaxed or not relaxed) are very similar 
to the bulk distribution (which is a broad Gaussian centered around 
six-fold rings \cite{jpcm}) and are therefore not reported in figure~4a.
The situation is different at the surfaces where the global trend after 
the annealing period is a decrease of the probability for a Si atom to be a 
member of a large ring (as expected, since a lot of bonds are broken) 
together with an increase of the probability to be a member of a small ring, 
more precisely, two-fold, three-fold and four-fold rings. A particular case 
of this behavior is the inversion of the proportion of five-fold and 
six-fold rings at the annealed surface. Energetically, there is no 
significant difference between these two ring sizes, therefore if six-fold 
rings are broken at the surface (because they are 
relatively large) a fraction of these rings will become five-fold  and
five-fold rings will become predominant: in a sense this corresponds to the 
reconstruction of an amorphous surface. An other significant feature is the 
relatively large proportion of two-fold rings at the annealed surface whereas 
it stays very weak in the bulk and at the freshly created surface. More 
precisely we find an average density of 0.25 rings per nm$^2$ which is in 
the range of the experimental values which vary between 0.1 and 0.4 rings 
per nm$^2$ \cite{watex2,dubois,infra1}. This proportion is smaller than the 
one found in ref \cite{wk} (0.6 rings per nm$^2$), where the statistics were 
done at 3000K and for silica clusters at equilibrium (which is certainly not
the case here). \\
Together with the ring size distribution we have also investigated the 
ring orientations by computing $<\cos^2\theta>$, for a given ring size $n$,
 within a given region of the sample, where $\theta$ is the angle 
between the normal to the surface and to the ring. The results for the 
surface region are reported in the bottom of figure 4 as a function of the
annealing time, and compared with those obtained in a 3D sample (which are 
similar to those in the interior region). 
In the bulk, all the $n$-fold rings are oriented in an 
isotropic way since the value of $<\cos ^2\theta>$ is close to $\frac{1}{3}$.
In the freshly created surface region, the $2,3$ and $4$-fold have still
an isotropic orientation, while the larger rings are oriented rather
parallel to the surface since $<\cos ^2\theta>$ is slightly larger than
$\frac{1}{3}$.
In the annealed surface layer we note that the rings larger than the 
three-fold ones have a strong tendency to orient parallel to the surface. 
On the contrary the two-fold rings are now perpendicular to the surface since 
$<\cos ^2\theta>$ is much smaller than $\frac{1}{3}$. This is a result of
the annealing and can therefore be extrapolated to a film in equilibrium
in which two-fold rings are certainly oriented perpendicular to the surface.
This is coherent with the {\em ab initio} study of Ceresoli et {\it al.} 
\cite{bernas} even if they had only one two-fold ring at the surface.\\
After the description of the structure let us now turn in the final part of
this work to the diffusion mechanism inside our thin silica films. The 
motivation of this study is connected to the idea that similarly to 
crystalline faces which develop a microscopic liquid film \cite{tosatti} the
surface layers of an amorphous sample show an enhanced diffusivity compared 
to the bulk \cite{litton,llayer}. The simplest way to address this question
is to compute the mean-squared displacement (MSD) $\langle r^2(t) \rangle$ = 
$\langle |{\bf r}_i(t)-{\bf r}_i(0)|^2 \rangle$ for a given particle type $i$.
This is what is represented in figure~5 for the oxygen atoms (the behavior of 
the silicon atoms is qualitatively similar as shown in Ref. \cite{wk}). The
interior and surface regions have been determined with respect to the
{\em initial} position of the atoms. In figure 5 we have also distinguished
the behavior of the non-bridging oxygens (NBOs) from that of the bridging 
oxygens (BOs). At the surface we find approximately 15 \% of NBOs while
in the interior only 0.5 \%  of the oxygen atoms are single-coordinated, 
which is in good agreement with a recent MD study of nanoporous silica 
\cite{leeuw}, especially for the number of NBOs in the surface region.
We have also calculated the components of the MSD that are parallel (
$1.5\times\langle x^2 + y^2 \rangle$) and perpendicular 
($3\times\langle z^2 \rangle$) to the surface, and these components are 
also represented in figure 5. In order to study the influence of
the annealing time, the MSD has been calculated for 42ps at the beginning
of the relaxation process (thin lines) and after an annealing time of 
105ps (bold lines).  
For the BO there is no difference in the MSD between the interior and the
surface region. After the annealing period, the displacement
of the atoms is smaller and more isotropic. Hence for a fully equilibrated
film the dynamics of the particles inside the sample is expected to 
be almost isotropic.\\
For the NBOs the situation is different. Firstly as expected, the NBOs 
diffuse more than the BOs. Next, after annealing,
the dynamics appear to be more {\em anisotropic} in both regions (interior
and surface). In the interior region the perpendicular MSD is more
important than the parallel one as if the NBOs were attracted towards the
surface. On the contrary in the surface layer the NBOs move parallel to the 
surface rather than perpendicular to it. This can be understood since most
of the Si-O bonds point outward of the surface and since it is easier for
an NBO to have a bond ``bending'' motion rather than a bond ``stretching''
motion. Therefore concerning the NBOs, we find that their dynamics is
anisotropic both in the interior and the surface region. Since this 
anisotropy is enhanced after annealing one can expect that this behavior
is also true in an equilibrated sample.
  
\section{conclusion}
We have used classical molecular dynamics to study the structural properties 
and the mean squared displacement of the particles inside an amorphous silica 
film. The first step in this work was to adapt our 
computational method to samples without periodic conditions in one 
spatial direction. This has been done using an original method which is 
fast and relatively accurate. With this technique we have calculated the 
density profile along the normal direction $z$ to the surface. 
We have found that the system is contracting since the equilibrium density
of the potential is greater than the mass density of the initial amorphous 
silica sample. This relaxation is rather fast and creating a surface seems
therefore to be a physical and efficient way to obtain the equilibrium
density of a given system. This shows also that the BKS potential
describes properly the physics inherent to the creation of free surfaces.
Like in liquid metals, the density profile exhibits a layering behavior 
that seems to be enhanced at the surface. This feature was interpreted 
as a sign of an increase of the short range order near the surface. The 
asymmetric peak near the surface is a sign of a stoichiometry breaking: 
while the silicon atoms are in excess in the inner part of the surface
the oxygen atoms are in excess in its outer part as already mentioned
in previous studies \cite{garofalini,wk}. We have calculated also the radial
pair distribution functions and the angle distributions and we found
the signature of the presence of 2-fold rings at the surface similarly 
to previous studies \cite{wk}. Therefore we have also focused our attention 
on the ring size distribution and ring orientation along the $z$-direction. 
We see that at the free surfaces, the most energetically unfavorable 
configurations (small rings) are enhanced and correlatively, the large rings 
show a tendency to disappear. A particular case of this feature is the 
inversion of the proportion of five-fold and six-fold rings. Also the 
orientation of the small-sized rings with respect to the surface 
(2-fold rings are perpendicular to the surface) agrees well with a recent 
{\em ab initio} study \cite{bernas}.
Concerning the MSD of the oxygen atoms, we find that the dynamics of
the BOs is unaffected by the presence of the surface. On the contrary the
dynamics of the NBOs is clearly anisotropic: in the interior (far from the 
surface) they move rather perpendicular to the surface while in the surface
region they move parallel to it.

{\bf \hrule}


\vskip 0.5cm
\begin{figure}
\psfig{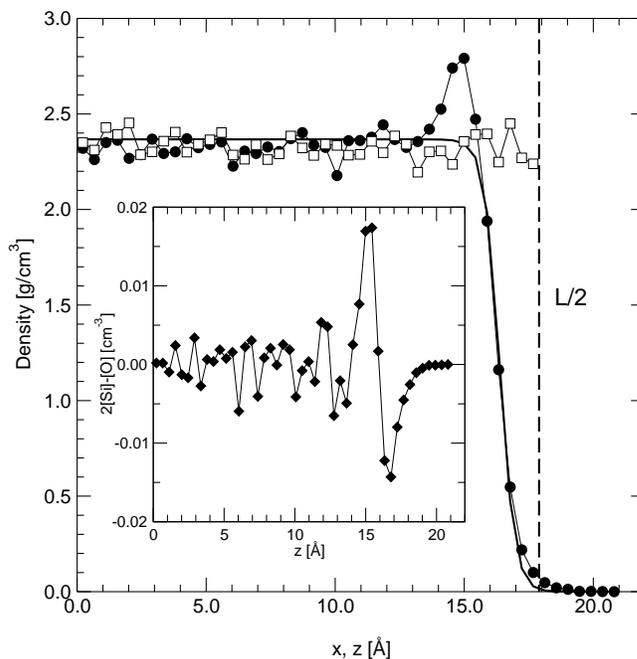}
\caption{
Density profile $\rho(z)$ in the direction $z$, perpendicular to the surface, from $z=0$ (median cut of the simulation box) up to the outer part 
of the surface. The filled circles correspond to the density calculated 
within slabs of width $\Delta z = 0.4475$ \AA\ for a sample size of $L=35.8$ 
\AA. The densities for slabs of opposite coordinates $z$ have been averaged.
The curve represents the fit of $\rho(z)$ using eq. (5).
In the inset the quantity $\Delta n = $  2[Si] - [O] is plotted 
as a function of $z$, where  [Si] and [O] are the numbers of Si and O atoms 
per unit volume calculated within the slabs.  
}
\label{Fig1}
\end{figure}
\vspace*{-2.5cm}
\begin{figure}
\psfig{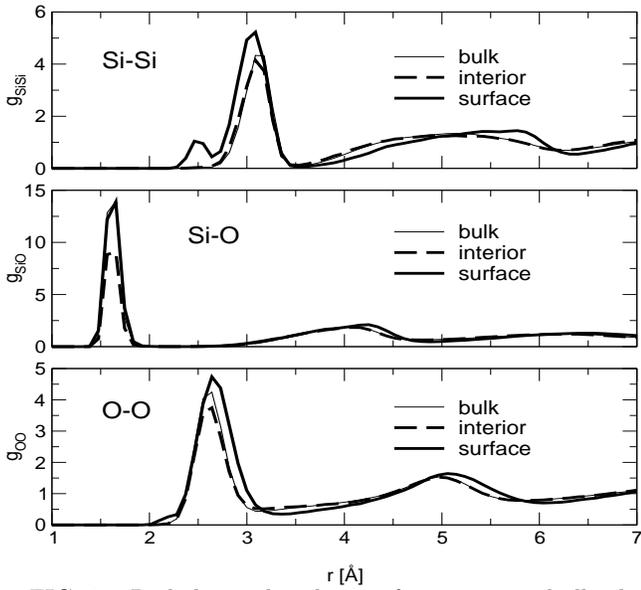}
\caption{
Radial pair distribution functions in a bulk silica sample 
(thin solid), in the interior region of our silica film (dashed) and in the 
surface region (solid bold) (see text for the definition of the regions).
}
\label{Fig2}
\end{figure}

\begin{figure}
\psfig{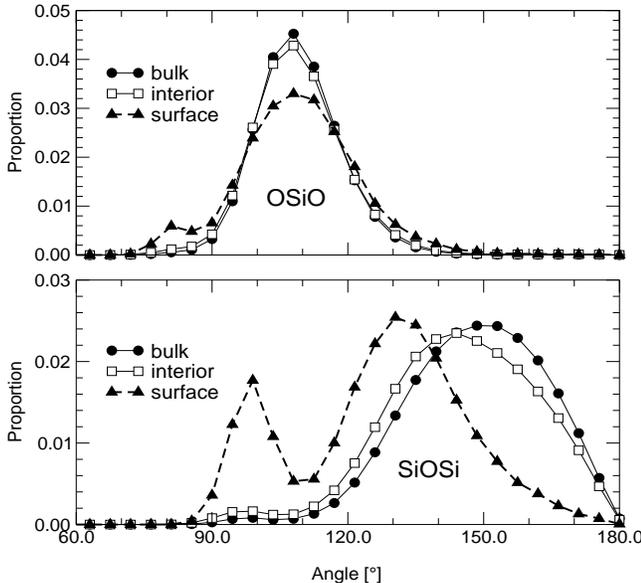}
\caption{
O-Si-O and Si-O-Si angle distributions in a bulk silica sample (circles), the interior (squares) and surface (diamonds) regions of our silica film. 
}
\label{Fig3}
\end{figure}

\vspace*{-2.5cm}
\begin{figure}
\psfig{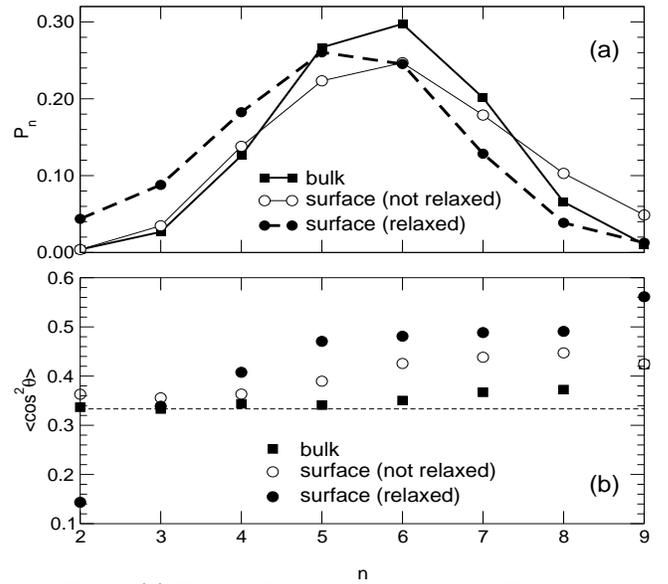}
\caption{
(a) Plot of $P_n$ versus $n$ in a bulk silica sample, 
in the initial surface region (not relaxed) and in a 42 ps annealed
surface region (relaxed) where $P_n$ is the probability that an atom 
in a given region is a member of an $n$-fold ring.\\
(b) Plot of $<\cos^2\theta>$ as a function of the ring size $n$, where $\theta$ is the angle between the direction perpendicular to the surface and the 
direction perpendicular to the ring. The dashed line represents 
$<\cos ^2\theta> = \frac{1}{3}$ corresponding to an isotropic orientation
of the rings.
}
\label{Fig4}
\end{figure}

\begin{figure}
\psfig{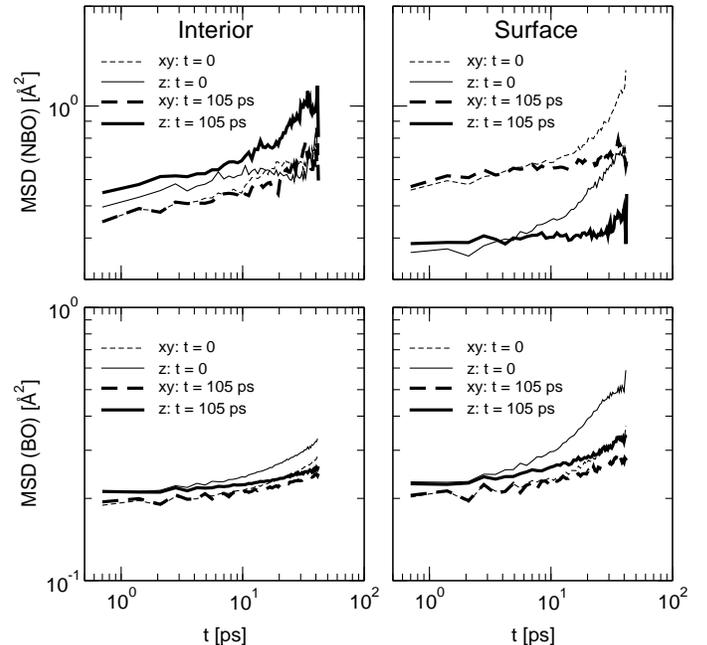}
\caption{
Plot of the mean-squared displacement (MSD) of the bridging 
oxygens (BO) and non-bridging oxygens (NBO) in the interior and surface 
regions as a function of the annealing time. We have distinguished the 
parallel component of the MSD $1.5\times\langle x^2 + y^2 \rangle$ (dashed
and labeled xy) and the perpendicular component $3\times\langle z^2 \rangle$ 
(solid and labeled z).
}
\label{Fig5}
\end{figure}


\end{document}